\documentclass[aps,pre,twocolumn,superscriptaddress,showpacs,preprintnumbers]{revtex4-1}%
\usepackage{natbib}
\usepackage[english]{babel}
\usepackage[dvips]{graphics}
\usepackage{graphicx,epsfig}
\usepackage{amsmath}
\usepackage{amsfonts}
\usepackage{amssymb}
\usepackage{xcolor}
\usepackage{multirow}
\usepackage[normalem]{ulem}
\usepackage{booktabs}
\usepackage{lineno}
\usepackage{url}
\usepackage{breakurl}
\usepackage[breaklinks]{hyperref}
\usepackage{graphicx}%
\usepackage{amsthm}

\usepackage{tikz}
\usepackage{pgfplots}
\pgfplotsset{compat=1.18}
\usepgfplotslibrary{fillbetween}

\def\tsc#1{\csdef{#1}{\textsc{\lowercase{#1}}\xspace}}
\tsc{WGM}
\tsc{QE}

\newcommand{\dn}{\textnormal{dn}}

\newcommand{\Hor}{\mathcal{\tilde H}}
\newcommand{\tor}{\tilde t}
\newcommand{\por}{\tilde p}
\newcommand{\Tor}{\tilde T}

\newcommand{\Hs}{\mathcal{H}}
\newcommand{\ts}{t}
\newcommand{\ps}{p}
\newcommand{\Ts}{T}

\begin{document}
\let\WriteBookmarks\relax
\def\floatpagepagefraction{1}
\def\textpagefraction{.001}


\title{Finite-time effects in periodically kicked systems}  

\author{F. Revuelta}
\affiliation{Grupo de Sistemas Complejos, 
   Escuela T\'ecnica Superior de Ingenier\'ia Agron\'omica, 
   Alimentaria y de Biosistemas,
   Universidad Polit\'ecnica de Madrid,
   Avenida Puerta de Hierro 2-4, 28040 Madrid, Spain.} 
\author{R. Chac\'{o}n}
\affiliation{Departamento de F\'{\i}sica Aplicada, E.I.I., Universidad de
                       Extremadura, Apartado Postal 382, E-06006 Badajoz, Spain}
\affiliation{Instituto de Computaci\'{o}n Cient\'{\i}fica Avanzada (ICCAEx),
                       Universidad de Extremadura, E-06006 Badajoz, Spain}
\author{F. Borondo}
\affiliation{Departamento de Qu\'imica, 
   Universidad Aut\'onoma de Madrid, 
   Cantoblanco, 28049 Madrid, Spain.}

\begin{abstract}
In this work, we study finite-time effects in ultracold atomic systems by considering time-dependent modulations with variable waveforms and durations. These two characteristics can be controlled by adjusting only a single parameter. For arbitrarily short pulses, our model recovers the paradigmatic kicked rotor while maintaining the impulse transmitted per period and unit amplitude constant. Furthermore, we demonstrate that finite-time effects have a profound impact on dynamical localization, a result that cannot be captured by the $\delta$-kicked-rotor model. Through a detailed analysis of the effects of different modulation amplitudes, periods, and waveforms, we identify the conditions for which dynamical localization is significantly enhanced. We show that the strength of dynamical localization increases sharply as the system approaches the $\delta$-kicked-rotor limiting case. Moreover, we establish the existence of an optimal value of the period that maximizes dynamical localization for given values of the amplitude and shape parameter.
\end{abstract}

\maketitle

\section{Introduction} \label{sec.intro}
Periodically-driven systems have been an active field of research activity over the last decades. The pioneering work of Kapitza showed the potential ability of a periodic driving to stabilize an inverted pendulum~\cite{Kapitza51, Haar65}. Ever since, external periodic modulations have been successfully applied to control and manipulate a large variety of systems, both from the classical and quantum points of view, such as the control of activated processes~\cite{Jung93} in classical metastable systems. Similarly, periodic electric fields are routinely used to confine ions in radio-frequency Paul traps~\cite{Paul90}. Periodic pulses are also commonly used in Raman spectroscopy~\cite{Smith19}, as well as in other processes of atomic ionization~\cite{Mainfray91, Paul01}. They have been also employed to generate exotic states of matter that do not show up in steady states, such as topological insulators~\cite{Goldman14, Ghosh20} and time-crystals~\cite{Else16}. Furthermore, periodic driving can enable other interesting phenomena otherwise not allowed. That is the case of the many-body localization~\cite{Das10, DAlessio13, Abanin19}, many-body scarring~\cite{Turner18, Serbyn21, Haldar21}, and the superfluid-to-Mott-insulator transition~\cite{Eckardt05}. The coherent destruction of tunneling~\cite{Grossmann91, Miyake13, Grifoni98} is another interesting effect caused by periodic drivings.

Under certain circumstances, the quantum behavior of a system greatly differs from its classical counterpart~\cite{Stoeckmann99, Reichl04}. This is the case of dynamical localization (DL), which is the quantum suppression of chaotic diffusion under time-dependent drivings~\cite{Dunlap86, Casati79}. For sufficiently short times, quantum wave packets rapidly spread mimicking the fast diffusion that takes place classically. Nevertheless, after a certain amount of time, quantum diffusion gets frozen due to the balance between chaotic diffusion and coherence effects that are absent in the classical description of the system.
This phenomenon of DL has attracted a lot of attention~\cite{Dunlap86, Medhet20}, partly due to its importance to understand the correspondence between classical and quantum mechanics within the field of quantum chaos, and, in addition, because of its strong connection with Anderson localization~\cite{Fishman82, Grempel84}, 
i.\,e.,
the quantum suppression of classical diffusion in disordered potentials~\cite{Anderson58, Billy08, Abrahams10}.

The breakdown of DL under quasiperiodic 
(by 
terms with inconmensurable frequencies)~\cite{Abal02, Ringot00, Lignier05} and
aperiodic (by addition of noise)~\cite{Paul19, Steck00, Klappauf98} drivings have also been studied. In this 
respect, it has been recently shown that DL can appear under quasiperiodic modulations for certain values of the modulation’s amplitude and frequency~\cite{Tiwari24}, as well as of the modulation waveform~\cite{Revuelta24}.

Several theoretical studies have been based on the kicked rotor, which 
is a very popular model~\cite{Fishman89, Delande13, Santhanam21, d'Arcy01, Hamilton22} to describe periodically-driven systems. The kicked rotor assumes a periodic modulation formed by a series of infinitely short pulses \emph{(kicks)} of equal amplitude. From the classical perspective, the kicked rotor can be described by a discrete map (the Chirikov-Taylor map)~\cite{LL10, Casati90}, something that certainly facilitates the numerical simulations. The quantum version of the kicked rotor has been also extensively studied in order to shed light on the classical-quantum correspondence in the presence of chaos~\cite{Carlo05, Hensinger01, Sadgrove12}.

When sufficiently fast periodic modulations are applied, the kicked rotor can provide an accurate description of the system~\cite{Moore94, Ammann98, Lignier05}. However, when the particle dynamics is not sufficiently slow, the finite duration and the waveform of the external modulation play a crucial role, making the instantaneous description provided by the kicked rotor ineffective. Thus, the main purpose of this work is the introduction and study of a model that overcomes these two limitations. We consider modulation transitions between two well-known limiting cases as a single parameter (the shape parameter, $m$) is changed. Namely, our model can vary from an integrable pendulum, where no time-dependence is considered, to a kicked rotor, where the periodic pulsatile modulation acts instantaneously. In contrast to other finite-duration models reported in the literature~\cite{Chacon01, Abal02, Chacon09, Revuelta15, Revuelta18, Naeem26}, the modulation considered in this paper induces the same dynamical effects regardless of the shape parameter by maintaining the transmitted impulse constant. In this way, we do not introduce spurious effects that may obscure our conclusions.

The remaining of this work is organized as follows. First, we introduce in Sec.~\ref{sec:system} our model system under study by describing the distinct characteristics of the time-dependent modulation considered. Second, we briefly describe in Sec.~\ref{sec:method} the numerical scheme that has been applied in the performed simulations of the system, both classically and quantum mechanically. Third, the results and the corresponding discussion are reported in Sec.~\ref{sec:results}. Finally, we present some conclusions and outlook in Sec.~\ref{sec.conclusions}.

\section{System} \label{sec:system}
\begin{figure}[hbtp!]
\includegraphics[width=0.9\columnwidth]{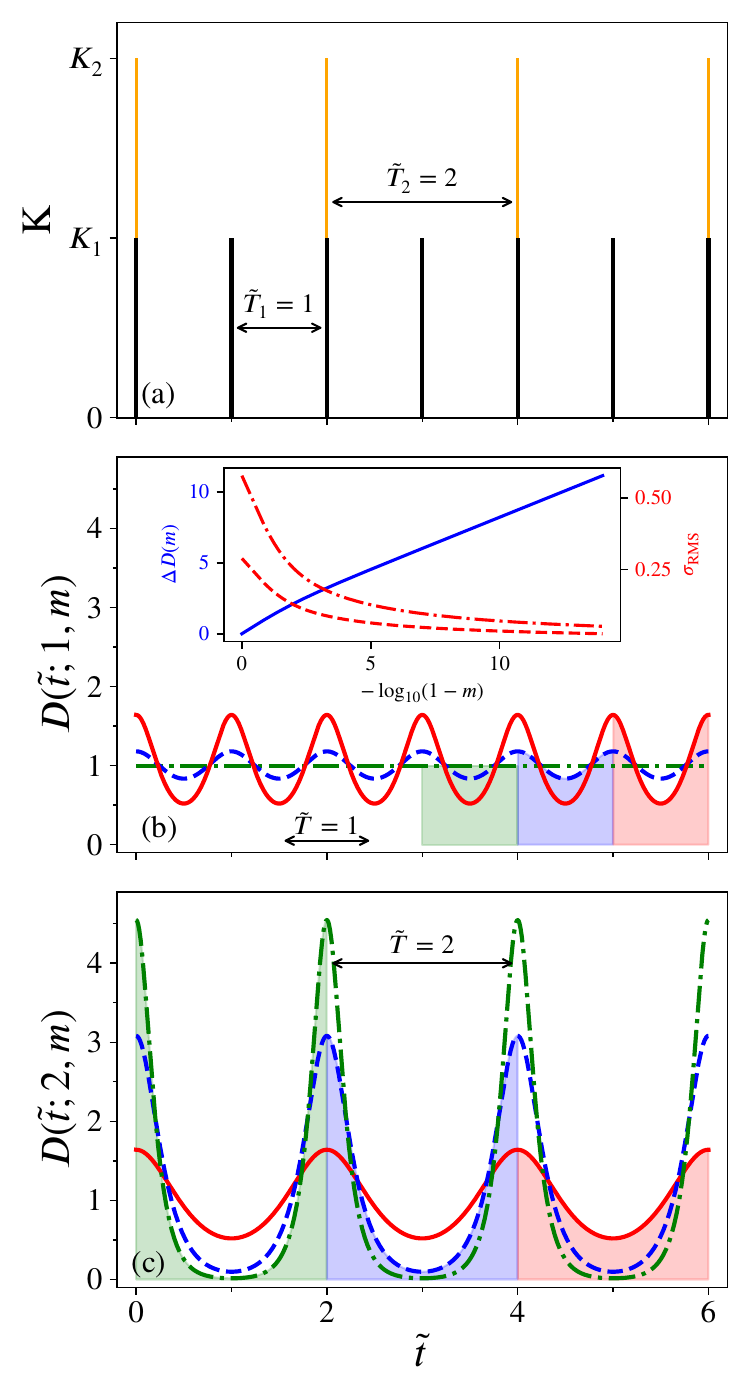}
\caption{Schematic illustration of the function~$D(\tor; 1, m)$ [cf. Eq.~\eqref{eq:D}]
                  in the limiting case~$m \rightarrow 1$ for two instances of amplitudes~$K_1$ ,$K_2$,
                 and periods~$T_1$, $T_2$, respectively, such that the impulse transmitted per unit period
                  is the same.
                   %
                  (b) Modulation~$D(\tor; 1, m)$ [cf. Eq.~\eqref{eq:D}] for
                   three values of the shape parameter:
                   $m=0$ (dashed-dotted green line),
                   $m=0.5$ (dashed blue line), and
                   $m=0.9$ (solid red line).
                   The shaded areas equal the impulse transmitted in each case
                   [cf. Eq.~\eqref{eq:Ior}].
                   Inset: Effective modulation amplitude~$\Delta D(m) = 
                   \max[D(\tor; \Tor, m)] - \min[D(\tor; \Tor, m)] 
$
                         (blue solid line),
                         and width~$\sigma_{\rm RMS}$ [cf. Eq.~\eqref{eq:sigma}]
                         of the pulses 
                         for~$\Tor=1$ (dashed red line) and~$\Tor=2$ (dashed-dotted red line)
                         as a function of
                         the shape parameter.
                   (c) Modulation~$D(\tor; 2, m)$ for
                   $m=0.9$ (solid red line),
                   $m=0.999$ (dashed blue line), and
                   $m=0.99999$ (dashed-dotted green line).
%
The colored areas in the same panels 
are equal because the larger
height in the peaks
associated with larger values of~$m$
is compensated by the lower
position of the minima
[cf. Eq.~\eqref{eq:Ior}].
%
%
%
}
\label{fig:1}
\end{figure}

The system under study consists of a non-interacting gas of ultracold atoms confined in an optical lattice with a trap depth that is subjected to a time-dependent modulation. Due to the lack of interaction between the particles, the atomic cloud can be described simply by considering the following Hamiltonian associated with a single unit-mass particle
\begin{eqnarray} \label{eq:H1}
   \Hor( \por,  x, \tor) = \frac{ \por^2}{2} +  F D( \tor; \Tor, m) ( 1 - \cos x),
\end{eqnarray}
where $\por$ is the momentum, $x$ is the position, $\tor$ is the time, and $F$ and $\Tor$ are the amplitude parameter and the period of the modulation, respectively. Function $D$~\cite{Chacon97} is given by
\begin{equation} \label{eq:D}
   D(\tor; \Tor, m) = \frac{2 K(m)}{\pi} \dn\left[  \frac{2 K(m) \tor}{\Tor}; m  \right], 
\end{equation}
with $K(m)$ being the complete elliptic integral of first kind
and~$\dn(u; m)$ the delta amplitude Jacobi elliptic function~\cite{Abramowitz65}. Notice that the effective amplitude of this modulation $\Delta D(m) = \max[D(\tor; \Tor, m)] - \min[D(\tor; \Tor, m)] = D(0; \Tor, m) - D(\Tor/2; \Tor, m)$ equals to $F D(0; \Tor, m)$, when $m \rightarrow 1$ [cf. inset of Fig.~\ref{fig:1}(b)].

Figure~\ref{fig:1} shows time series of the modulation of Eq.~\eqref{eq:D}. As illustrated in Fig.~\ref{fig:1}(a), the periodic $\delta$ function $\delta_1 (\tor; \Tor) \equiv \lim_{m \rightarrow 1}  D(\tor; \Tor, m)$ can be approximated by changing the amplitude $K_i$ and the period $\Tor_i$. By increasing the period of the function $D$ [Eq.~\eqref{eq:D}] results in series of wider peaks [compare Figs.~\ref{fig:1}(b) and~\ref{fig:1}(c)].

Remarkably, the shape parameter $m$ offers an additional degree of freedom to fine-tune the waveform of $D(\tor; \Tor, m)$. For $m = 0$ [see dashed-dotted green line in Fig.~\ref{fig:1}(b)], a constant force of amplitude 1 is obtained, since $\lim_{m \rightarrow 0} D(\tor; \Tor, m) = 1$. In this scenario, no pulsatile modulation is applied and the depth of the optical lattice remains constant over time. Consequently, no DL can be observed due to the lack of a time-periodic driving.

Figures~\ref{fig:1}(b) and~\ref{fig:1}(c) show that the value of $D(\tor; \Tor, m)$ increases with $m$ at $\tor = 0, \Tor, 2\Tor, \ldots$, and decreases at $\tor = \Tor/2, 3\Tor/2, \ldots$. As a consequence, a series of taller and narrower peaks emerges as $m$ is increased. In fact, as depicted in the inset of Fig.~\ref{fig:1}(b), the height of the pulses depends exponentially on the value $m$.

Conversely, the width of the pulses, defined as the root mean square~\cite{Agrawal2010}
\begin{eqnarray} \label{eq:sigma}
    \sigma_{\rm RMS} = \sqrt{ \frac{\int_{-\Tor/2}^{\Tor/2} \tor^2 \vert D(\tor; \Tor, m ) \vert^2 d\tor}{\int_{-\Tor/2}^{\Tor/2} \vert D(\tor; \Tor, m ) \vert^2 d\tor} },
\end{eqnarray}
reduces with $m$. Also, notice in the red curves of the inset of Fig.~\ref{fig:1}(b) that the longer the period, the larger the width. Notably, the limit $m \rightarrow 1$ in Eq.~\eqref{eq:H1} yields a kicked rotor of period $\Tor$ with infinite amplitude as $\lim_{m \rightarrow 1} D(\tor; \Tor, m) = \sum_{n=0}^{\infty}  \delta(n\Tor)$ with $\delta_n(x)$ being the Dirac delta.

An interesting property of Eq.~\eqref{eq:D} is that the impulse transmitted per unit of amplitude over one period of time $\Tor$ does not depend on the shape parameter $m$ but only on the period considered,
\begin{equation} \label{eq:Ior}
    I(\Tor) = \int_0^{\Tor}  D(\tor; \Tor, m) dt = \Tor.
\end{equation}
Thus, model~\eqref{eq:H1} allows to include finite-time effects while maintaining the impulse transmitted per unit period and unit amplitude constant.

Note that the kinetic energy variation after one period does not depend on the shape parameter $m$ but only on the net impulse, $F \tilde T$, transmitted by the periodic modulation
\begin{eqnarray}
&&   E_k \left[ (n+1) \Tor \right] - E_k \left[ n \Tor \right] = \nonumber \\
&=& - \dot x(\tau) \sin \left[ x(\tau) \right] \int_{n\Tor}^{(n+1)\Tor} F D(\tor; \Tor, m) d\tor \nonumber \\
& = & - \dot x(\tau) \sin \left[ x(\tau) \right] F \Tor ,
\end{eqnarray}
where we use Eq.~\eqref{eq:Ior}, and $\tau \in [n \Tor, (n+1) \Tor]$.

Moreover, the width of the stochastic separatrix layer associated with the classical dynamics of Hamiltonian~\eqref{eq:H1} can be analytically obtained by performing a simple scaling in the modulation amplitude $F$ in the Melnikov analysis~\cite{LL10, Guckenheimer02} reported in Ref.~\cite{Revuelta18}. This width serves as a practical and effective tool for identifying the onset of classical diffusion:
\begin{align} \label{eq:d}
d(\Tor, F, m) \approx\frac{4\pi^{3}}{\Tor^{2}}\sum_{n=1}^{\infty}n^{2} c_{n}\left(m\right) b_{n}\left( \Tor, F, m \right) ,
\end{align}
where
\begin{eqnarray}
b_{n}\left( \Tor,F, m\right) &\equiv& \operatorname{csch}\left( \frac{n\pi^{2} \sqrt \pi}{\Tor\sqrt{2 F K\left( m\right) (1-\sqrt{1-m})}}\right) ,\nonumber\\
c_{n}\left( m\right) &\equiv&  \frac{\pi\operatorname{sech}\left[ n\pi K\left( 1-m\right) / K\left( m\right) \right] }{\left( 1-\sqrt{1-m}\right) K(m)}.
\end{eqnarray}

\begin{figure}[hbtp!]
\includegraphics[width=0.97\columnwidth]{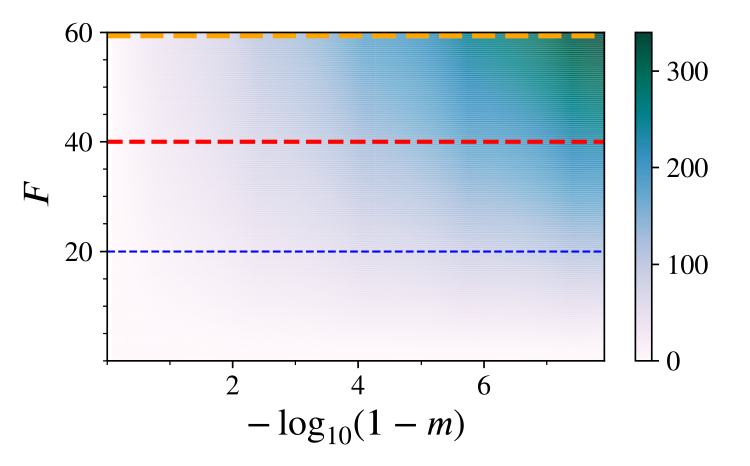}
\caption{Chaotic layer width in the $F-m$ plane 
[cf. Eq.~\eqref{eq:d}]
for $\Tor=1$.
Sections of the 
parameter chaotic layer width 
corresponding to the values indicated by 
the dashed  lines ($F=20, 40$, $60$) are
shown in Fig.~\ref{fig:F-m_T1}(d).
}
\label{fig:d_F-m_T1}
\end{figure}

Figure~\ref{fig:d_F-m_T1} shows the width of the chaotic layer [Eq.~\eqref{eq:d}] as a function of $F$ and $m$ for $\Tor=1$. For sufficiently low values of $F$, the motion remains predominantly regular, resulting in a narrow chaotic layer. The width is strictly zero for $m=0$. In contrast, as $F$ or $m$ increase, the width of the chaotic layer expands, indicating the expansion of chaos in phase space.


\section{Numerical simulations} \label{sec:method}
This section describes the numerical schemes applied in our simulations. First, we scaled the time in Eq.~\eqref{eq:H1} as $\ts = \tor \sqrt{ 2 K(m) / \pi }$, transforming Hamiltonian~\eqref{eq:H1} into
\begin{eqnarray} \Hs( \ps, x, \ts) = \frac{ p^2}{2 } + F \dn\left[ \sqrt{2 K(m)\pi} \frac{\ts}{\Tor}; m \right] ( 1 - \cos x ) , \label{eq:H}
\end{eqnarray}
where $\ps = \por \sqrt{\pi / [ 2 K(m)] }$ is the scaled momentum, and energy is measured in $2 K(m)/\pi$ units. Note that the new unit of time depends on the shape parameter $m$. The modulation considered in Eq.~\eqref{eq:H} has a period $\Ts = \Tor \sqrt{2 K(m) / \pi }$. The impulse transmitted per unit of amplitude over one period $\Ts$ is given by
\begin{equation} \label{eq:Is}
    I(\Ts) = \int_0^{\Ts} \dn \left[ \sqrt{2 K(m)\pi} \frac{\ts}{\Tor}; m \right] = \Ts,
\end{equation}
which solely depends on the driving period $\Ts$ and not on $m$.

Second, we calculated the normalized root-mean-squares of the classical $p_C$ and quantum $p_Q$ momenta by computing:
\begin{equation}
\Delta p_{i} = \sqrt{\left\langle p_{i}^{2}\right\rangle -\left\langle p_{i}\right\rangle ^{2}},\qquad i=\left\{ C,Q\right\} . \label{eq:Dp}
\end{equation}
In the presence of DL, chaotic diffusion is significantly suppressed in the quantum regimen, leading to $\Delta p_Q \ll \Delta p_C$. As a result, a substantial disparity arises between classical and quantum momentum dispersions, quantified by
\begin{equation}
\Delta p_{C-Q}=\Delta p_{C}-\Delta p_{Q} \label{eq:DpCQ}.
\end{equation}

The computation of $\Delta p_C$ relies on the numerical integration of the Liouville equation
\begin{equation} \label{eq:Liou}
\left[ \partial_t+p\partial_{x} - F \dn\left[ 2 K(m) \frac{t}{T}; m \right] \sin x \right] P\left(x,p;t\right)=0
\end{equation}
subject to periodic boundary conditions $P\left( x+L,p;t\right) =P\left( x,p;t\right) $, where $L=2n\pi$ denotes the size of the quantization box ($n\in\mathbb{Z}^{+}$). The classical momentum distribution is obtained as
\begin{equation}
P_{C}\left( p,t\right) =\int P\left( x,p;t\right) dx,
\end{equation}
where $P\left( x,p;t\right) $ represents the evolution of $2\times10^{4}$ trajectories initialized over a Gaussian distribution with width $\Delta p_{0}=0.386$ up to 320 time units using a fourth-order Runge-Kutta method~\cite{NR07}.

On the other hand, $\Delta p_Q$ is calculated with the fast Fourier transform method~\cite{NR07} by solving the time-dependent Schrödinger equation:
\begin{eqnarray}\label{eq:Sch}
i\hbar_{\rm{eff}}\partial_t\psi (x, t)= \left[ - \frac{\hbar_{\rm{eff}}^2}{2} \partial_{xx}^{2} +F \dn\left[ 2 K(m) \frac{t}{T}; m \right] (1- \cos x) \right] \psi (x, t), \quad
\end{eqnarray}
for an initial Gaussian wave packet
\begin{equation}
\psi\left( x,0\right) =\frac{1}{(\pi\Delta x_{0})^{1/4}}\exp\left[ -\frac{\left( x-x_{0}\right) ^{2}}{2\Delta x_{0}}+\frac{ixp_{0}} {\hbar_{\mathrm{eff}}}\right] , \label{eq:gauss}
\end{equation}
centered at $x_{0}\in\left[ 0,L\right] $ with $p_{0}=0$ and $\Delta x_{0}=\hbar_{\mathrm{eff}}/\Delta p_{0}$. Here, $\hbar_{\mathrm{eff}} = \hbar \sqrt{2 \pi / K(m)} 2 \pi^2 / ( M a^2) $ is the effective Planck constant. Quantum distributions $P_{Q}\left(p,t\right) $ were obtained by averaging wave packet time evolutions up to 320 time units.

\section{Results and discussion} \label{sec:results}
\begin{figure}[hbtp!]
\includegraphics[width=\columnwidth]{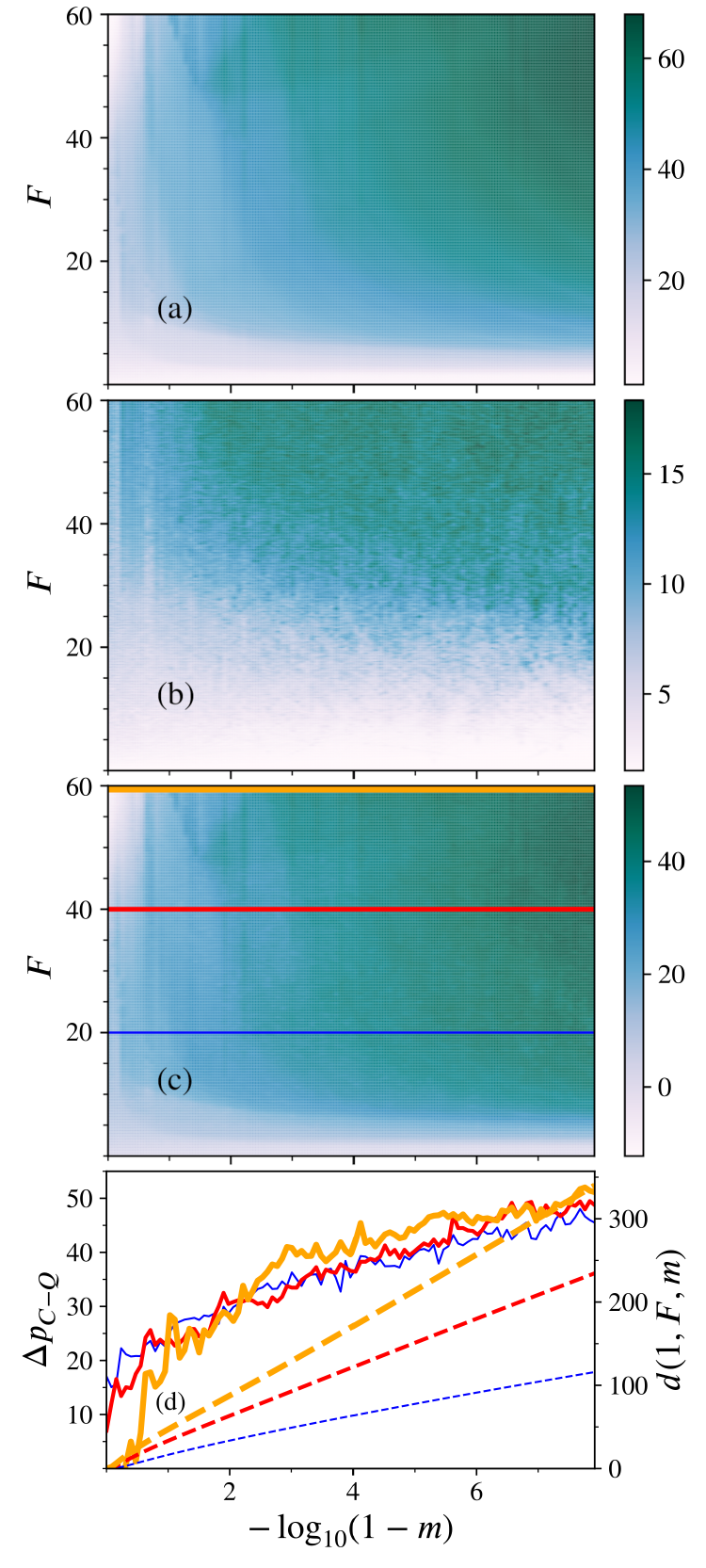}
\caption{Comparison between the strengths of 
                  dynamical localization 
                 and chaos for~$\Ts=1$
                  in the parameter plane~$m-F$.
                 (a) Classical momentum dispersion~$\Delta p_C$,
                 (b) quantum momentum dispersion~$\Delta p_Q$
                 given by Eq.~\eqref{eq:D},
                 and (c) their difference~$\Delta p_{C-Q}=\Delta p_C-\Delta p_Q$
                 [Eq.~\eqref{eq:DpCQ}].
                 (d) Sections of~$\Delta p_{C-Q}$ 
                 for three
                 values of the modulation amplitude:
                 $F=20, 40$, $60$ 
                [solid lines, also shown in (c)]
                 and chaotic layer width [Eq.~\eqref{eq:d}]
                [dashed lines, also shown as in Fig.~\ref{fig:d_F-m_T1}].
}
\label{fig:F-m_T1}
\end{figure}

Figure~\ref{fig:F-m_T1}(a) shows the classical dispersion $\Delta p_C$ [Eq.~\eqref{eq:Dp}] as a function of amplitude $F$ and shape parameter $m$. Dispersion generally increases with $F$, while the modulation waveform strongly activates chaotic diffusion. For small $m$, chaotic motion is weak, yielding small $\Delta p_C$. In the limit $m \rightarrow 1$, $\Delta p_C$ increases dramatically due to the divergence in effective modulation amplitude [Fig.~\ref{fig:1}(b)].

A similar trend is observed in the quantum regime [Fig.~\ref{fig:F-m_T1}(b)], though quantum dispersion $\Delta p_Q$ is notably smaller. Figure~\ref{fig:F-m_T1}(c) displays the momentum difference $\Delta p_{C-Q}$ [Eq.~\eqref{eq:DpCQ}], where large values signify DL. Figure~\ref{fig:F-m_T1}(d) shows sections of $\Delta p_{C-Q}$ for three values of $F$. For $m=0$, $\Delta p_{C-Q} \to 0$ and no DL occurs, whereas for large $m$, DL is strongly enabled.

\begin{figure}[hbtp!]
\includegraphics[width=\columnwidth]{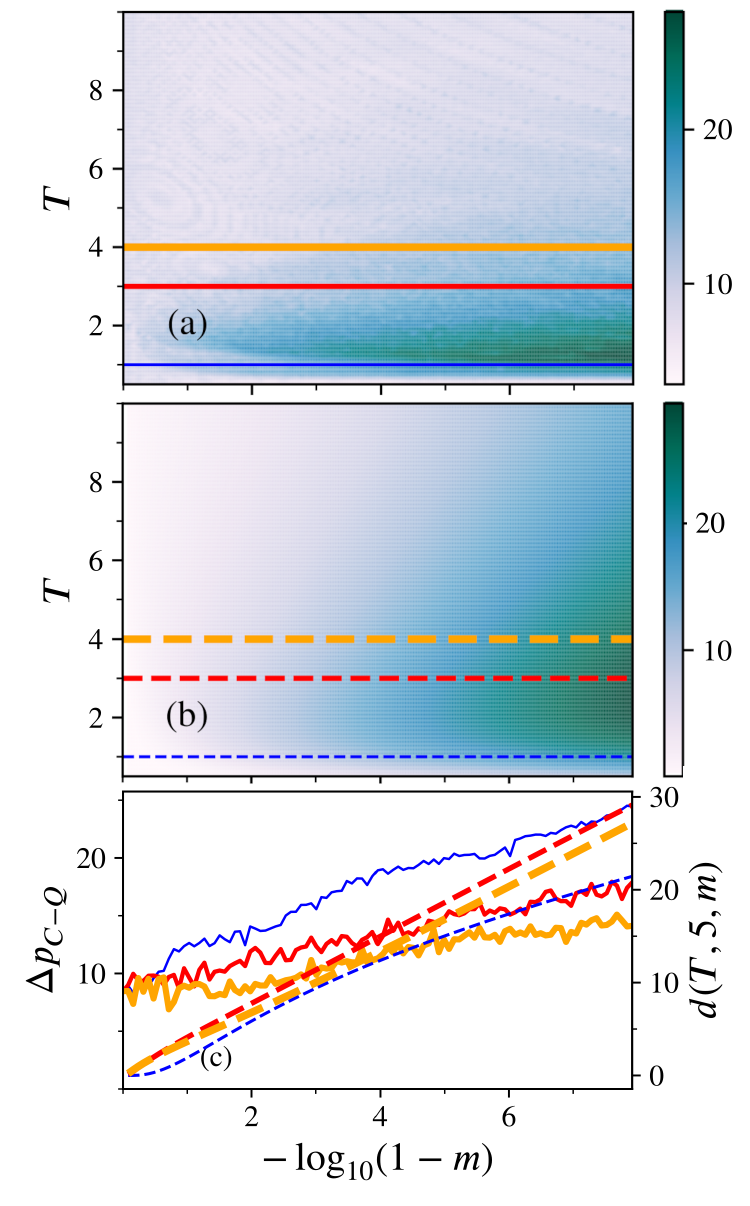}
\caption{Comparison between the strengths of 
                  dynamical localization 
                 and chaos for~$F=5$
                   in the parameter plane~$m-\Ts$.
                  (a) Difference between the classical and quantum momentum
                  dispersions~$\Delta p_{C-Q}=\Delta p_C-\Delta p_Q$ [Eq.~\eqref{eq:DpCQ}].
                  (b) Chaotic layer width [Eq.~\eqref{eq:d}].
                 (c) $\Delta p_{C-Q}$ vs $m$
                 for: 
                $T=1$ (blue), $3$ (red),  and~$4$ (orange)
                [solid lines, also shown in (a)],
                 and chaotic layer width [Eq.~\eqref{eq:d}]
                [dashed lines, also shown in~(b)].
}
\label{fig:T-m_F5}
\end{figure}

Figures~\ref{fig:T-m_F5}(a) and \ref{fig:T-m_F5}(b) show $\Delta p_{C-Q}$ and $d(T, F=5, m)$ in the $m-T$ plane, with constant-$T$ sections in Fig.~\ref{fig:T-m_F5}(c). Both quantities increase with $m$ even while keeping net impulse constant.

\begin{figure}[hbtp!]
\includegraphics[width=0.95\columnwidth]{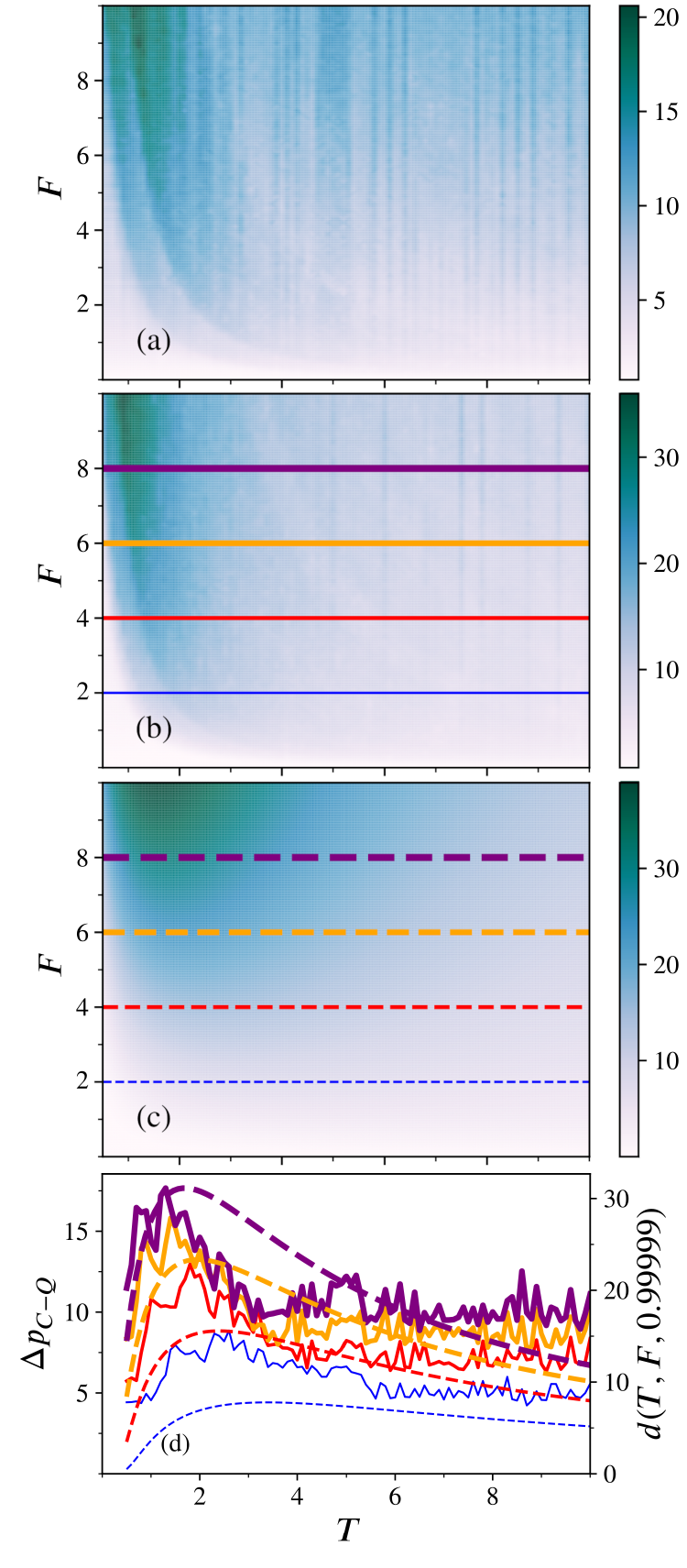}
\caption{Comparison between the strengths of 
                  dynamical localization 
                  and chaos 
                   in the parameter plane~$\Ts-F$.
                  The difference between the classical and quantum momentum
                  dispersions~$\Delta p_{C-Q}=\Delta p_C-\Delta p_Q$
                 corresponds to (a) $m=0.9$,
                  and (b) $m=0.99999$.
                  (c) Chaotic layer width [Eq.~\eqref{eq:d}] for $m=0.99999$.
                  (d) Difference~$\Delta p_{C-Q}$ for $m=0.99999$
                        and chaotic layer width
                         as a function of the period for four values of the amplitude
                        [cf. vertical lines in (b) and (c)].
}
\label{fig:T-F}
\end{figure}

Figure~\ref{fig:T-F} shows DL strength in the $F-T$ parameter plane for $m=0.9$ and $m=0.99999$. DL strength is visibly greater for $m=0.99999$. Crucially, $\Delta p_{C-Q}$ presents a non-monotonic behavior with $T$, exhibiting a clear maximum. This demonstrates the existence of an optimal modulation period that maximizes DL.


\section{Conclusions and Outlook} \label{sec.conclusions}
In this work, we have studied finite-time effects on dynamical localization in an ultracold atomic gas in a periodically modulated optical lattice. The chosen modulation transitions continuously from a static lattice ($m=0$) to a periodic $\delta$-kick driving ($m \to 1$), 
and more importantly,
while keeping the transmitted impulse per unit period and amplitude rigorously constant.

We confirmed a strong correlation between the classical chaotic layer width and quantum-classical dispersion difference $\Delta p_{C-Q}$. We demonstrated that dynamical localization is significantly enhanced as $m \rightarrow 1$. This presents a notable result: despite the transmitted impulse remaining unchanged, the strength of chaos and DL increase sharply because the modulation pulses become narrower and higher. Finally, we showed that an optimal period exists that maximizes dynamical localization for given amplitude and shape parameters.

\section*{Declaration of Competing Interest}
The authors declare that they have no known competing financial interests or personal relationships that could have appeared to influence the work reported in this paper.

\section*{Data Availability}
Data will be made available on reasonable request.

\section*{Acknowledgments}
This work has been partially supported by Grant PID2021-122711NB-C21 funded by MCIN/AEI/10.13039/501100011033 (F.B, F.R.), by the European Union's Horizon 2020 Research and Innovation Program under Marie Skłodowska-Curie Grant No. 734557 (F.B., F.R.), and by Comunidad de Madrid under Grant APOYO-JOVENES-4L2UB6-53-29443N (F.R.). RC acknowledges support from Junta de Extremadura (JEx) Project No. GR24101 cofinanced by FEDER funds.


\bibliography{delta_kicked_rotor}

@article{Abanin19,
  title = {Colloquium: Many-body localization, thermalization, and entanglement},
  author = {Abanin, Dmitry A. and Altman, Ehud and Bloch, Immanuel and Serbyn, Maksym},
  journal = {Rev. Mod. Phys.},
  volume = {91},
  issue = {2},
  pages = {021001},
  numpages = {26},
  year = {2019},
  month = {May},
  publisher = {American Physical Society},
  doi = {10.1103/RevModPhys.91.021001},
  url = {https://link.aps.org/doi/10.1103/RevModPhys.91.021001}
}

@book{Agrawal2010,
  author     = {Agrawal, Govind P.},
  title      = {Fiber-Optic Communication Systems},
  edition    = {4th},
  publisher = {John Wiley \& Sons},
  series     = {Wiley Series in Microwave and Optical Engineering},
  year       = {2010},
  isbn       = {978-0-470-50511-3}
}

@book{Abrahams10,
  editor    = {E. Abrahams},
  title     = {50 Years of Anderson Localization},
  publisher = {World Scientific},
  address   = {Singapore},
  year      = {2010},
}

@article{Carlo05,
  title = {Quantum Ratchets in Dissipative Chaotic Systems},
  author = {Carlo, Gabriel G. and Benenti, Giuliano and Casati, Giulio and Shepelyansky, Dima L.},
  journal = {Phys. Rev. Lett.},
  volume = {94},
  issue = {16},
  pages = {164101},
  numpages = {4},
  year = {2005},
  month = {Apr},
  publisher = {American Physical Society},
  doi = {10.1103/PhysRevLett.94.164101},
  url = {https://link.aps.org/doi/10.1103/PhysRevLett.94.164101}
}

@article{Casati90,
  author    = {Casati, G. and Guarneri, I. and Shepelyansky, D.},
  title     = {Anderson Transition in a One-Dimensional System with Three Incommensurate Frequencies},
  journal   = {Physica A},
  volume    = {163},
  pages     = {205--210},
  year      = {1990}
}

@article{Chacon97,
    author = {Chacón, Ricardo},
    title = {Chaos and geometrical resonance in the damped pendulum subjected to periodic pulses},
    journal = {Journal of Mathematical Physics},
    volume = {38},
    number = {3},
    pages = {1477-1483},
    year = {1997},
    month = {03},
    issn = {0022-2488},
    doi = {10.1063/1.531816}
}

@misc{Delande13,
  author    = {Delande, Dominique},
  title     = {Kicked Rotor and Anderson Localization},
  howpublished = {Boulder School on Condensed Matter Physics},
  year      = {2013},
  note      = {Lecture notes, Boulder School on Condensed Matter Physics, University of Colorado, Boulder}
}

@article{Else16,
  title = {Floquet Time Crystals},
  author = {Else, Dominic V. and Bauer, Bela and Nayak, Chetan},
  journal = {Phys. Rev. Lett.},
  volume = {117},
  issue = {9},
  pages = {090402},
  numpages = {5},
  year = {2016},
  month = {Aug},
  publisher = {American Physical Society},
  doi = {10.1103/PhysRevLett.117.090402},
  url = {https://link.aps.org/doi/10.1103/PhysRevLett.117.090402}
}

@article{Eckardt05,
  title = {Superfluid-Insulator Transition in a Periodically Driven Optical Lattice},
  author = {Eckardt, Andr\'e and Weiss, Christoph and Holthaus, Martin},
  journal = {Phys. Rev. Lett.},
  volume = {95},
  issue = {26},
  pages = {260404},
  numpages = {4},
  year = {2005},
  month = {Dec},
  publisher = {American Physical Society},
  doi = {10.1103/PhysRevLett.95.260404},
  url = {https://link.aps.org/doi/10.1103/PhysRevLett.95.260404}
}

@article{Goldman14,
  author    = {Goldman, N. and Dalibard, J.},
  title     = {Periodically Driven Quantum Systems: Effective Hamiltonians and Engineered Gauge Fields},
  journal   = {Phys. Rev. X},
  volume    = {4},
  number    = {3},
  pages     = {031027},
  year      = {2014},
  doi       = {10.1103/PhysRevX.4.031027},
}

@article{Ghosh20,
  title = {Higher order topological insulator via periodic driving},
  author = {Ghosh, Arnob Kumar and Paul, Ganesh C. and Saha, Arijit},
  journal = {Phys. Rev. B},
  volume = {101},
  issue = {23},
  pages = {235403},
  numpages = {9},
  year = {2020},
  month = {Jun},
  publisher = {American Physical Society},
  doi = {10.1103/PhysRevB.101.235403},
  url = {https://link.aps.org/doi/10.1103/PhysRevB.101.235403}
}

@article{Grifoni98,
title = {Driven quantum tunneling},
journal = {Phys. Rep.},
volume = {304},
number = {5},
pages = {229-354},
year = {1998},
issn = {0370-1573},
doi = {https://doi.org/10.1016/S0370-1573(98)00022-2},
url = {https://www.sciencedirect.com/science/article/pii/S0370157398000222},
author = {Grifoni, Milena and Hänggi, Peter}
}

@book{Haar65,
  editor    = {D. ter Haar},
  title     = {Collected Papers of P. L. Kapitza, Vol. II},
  publisher = {Pergamon Press},
  address   = {Oxford},
  year      = {1965},
}

@article{Hensinger01,
  title     = {Dynamical tunnelling of ultracold atoms},
  author    = {Hensinger, W. K. and H{\"a}ffner, H. and Browaeys, A. and Heckenberg, N. R. and Helmerson, K. and McKenzie, C. and Milburn, G. J. and Wilson, A. C. and Rubinsztein-Dunlop, H.},
  journal   = {Nature},
  volume    = {412},
  pages     = {52--55},
  year      = {2001},
  doi       = {10.1038/35083510},
  url       = {https://doi.org/10.1038/35083510},
}

@article{Jung93,
  author    = {Peter Jung},
  title     = {Periodically Driven Stochastic Systems},
  journal   = {Phys. Rep.},
  volume    = {234},
  number    = {4-5},
  pages     = {175--295},
  year      = {1993},
  publisher = {North-Holland},
  doi       = {10.1016/0370-1573(93)90035-J},
}

@article{Kapitza51,
  author    = {P. L. Kapitza},
  title     = {},
  journal   = {Zh. Eksp. Teor. Fiz.},
  volume    = {21},
  pages     = {588},
  year      = {1951},
}

@article{Klappauf98,
  title = {Observation of Noise and Dissipation Effects on Dynamical Localization},
  author = {Klappauf, B. G. and Oskay, W. H. and Steck, D. A. and Raizen, M. G.},
  journal = {Phys. Rev. Lett.},
  volume = {81},
  issue = {6},
  pages = {1203--1206},
  numpages = {0},
  year = {1998},
  month = {Aug},
  publisher = {American Physical Society},
  doi = {10.1103/PhysRevLett.81.1203},
  url = {https://link.aps.org/doi/10.1103/PhysRevLett.81.1203}
}

@article{Mainfray91,
doi = {10.1088/0034-4885/54/10/002},
url = {https://dx.doi.org/10.1088/0034-4885/54/10/002},
year = {1991},
month = {oct},
publisher = {},
volume = {54},
number = {10},
pages = {1333},
author = {Mainfray, G. and  Manus, G.},
title = {Multiphoton ionization of atoms},
journal = {Rep. Progr. Phys.}}

@article{Miyake13,
  author    = {Miyake, Hirokazu and Siviloglou,  Georgios A. and Kennedy, Colin J. and  Burton, William Cody and Ketterle,  Wolfgang},
  title     = {Realizing the Harper Hamiltonian with Laser-Assisted Tunneling in Optical Lattices},
  journal   = {Phys. Rev. Lett.},
  volume    = {111},
  pages     = {185302},
  year      = {2013},
  doi       = {10.1103/PhysRevLett.111.185302},
}

@article{Naeem26,
title = {Controlling dynamical localization by a non-ideal pulsatile function in a modulated gravitational cavity},
journal = {Chaos, Solitons \& Fractals},
volume = {202},
pages = {117471},
year = {2026},
issn = {0960-0779},
doi = {https://doi.org/10.1016/j.chaos.2025.117471},
author = {Muhammad Naeem and Muhammad Ayub and Amjad Sohail and Khalid Naseer}
}

@article{Paul01,
  title     = {Observation of a Train of Attosecond Pulses from High Harmonic Generation},
  author    = {Paul, P. M.  and  Toma, E. S. and Breger,  P. and Mullot,  G. and Auge\'e, F. and Balcou,  Ph. and Muller, H. G. and Agostini,  P.},
  journal   = {Science},
  volume    = {292},
  number    = {5522},
  pages     = {1689--1692},
  year      = {2001},
  doi       = {10.1126/science.1059413},
}

@article{Paul90,
  title = {Electromagnetic traps for charged and neutral particles},
  author = {Paul, Wolfgang},
  journal = {Rev. Mod. Phys.},
  volume = {62},
  issue = {3},
  pages = {531--540},
  numpages = {0},
  year = {1990},
  month = {Jul},
  publisher = {American Physical Society},
  doi = {10.1103/RevModPhys.62.531},
  url = {https://link.aps.org/doi/10.1103/RevModPhys.62.531}
}

@article{Paul19,
  title = {Nonmonotonic diffusion rates in an atom-optics L\'evy kicked rotor},
  author = {Paul, Sanku and Sarkar, Sumit and Vishwakarma, Chetan and Mangaonkar, Jay and Santhanam, M. S. and Rapol, Umakant},
  journal = {Phys. Rev. E},
  volume = {100},
  issue = {6},
  pages = {060201},
  numpages = {5},
  year = {2019},
  month = {Dec},
  publisher = {American Physical Society},
  doi = {10.1103/PhysRevE.100.060201},
  url = {https://link.aps.org/doi/10.1103/PhysRevE.100.060201}
}

@book{Reichl04,
  title     = {The Transition to Chaos: Conservative Classical Systems and Quantum Manifestations},
  author    = {Reichl, Linda E.},
  year      = {2004},
  edition   = {2nd},
  publisher = {Springer},
  address   = {New York},
  isbn      = {978-0387406153},
}

@article{Revuelta24,  title = {Dynamical localization in nonideal kicked rotors driven by two competing pulsatile modulations},
  author = {Revuelta, F. and Chac\'on, R. and Borondo, F.},
  journal = {Phys. Rev. E},
  volume = {110},
  issue = {5},
  pages = {054202},
  numpages = {10},
  year = {2024},
  month = {Nov},
  publisher = {American Physical Society},
  doi = {10.1103/PhysRevE.110.054202}
}

@article{Sadgrove12,
  author    = {Sadgrove, M. and Mullins, T. and Parkins, S. and Leonhardt, R.},
  title     = {Quantum Resonant Effects in the Delta-Kicked Rotor Revisited},
  journal   = {Eur. Phys. J. D},
  volume    = {66},
  pages     = {315},
  year      = {2012},
  doi       = {10.1140/epjd/e2012-30355-3}
}

@article{Santhanam21,
title = {Quantum kicked rotor and its variants: Chaos, localization and beyond},
journal = {Phys. Rep.},
volume = {956},
pages = {1-87},
year = {2022},
issn = {0370-1573},
doi = {https://doi.org/10.1016/j.physrep.2022.01.002},
url = {https://www.sciencedirect.com/science/article/pii/S0370157322000047},
author = {Santhanam, M.S.  and  Paul, Sanku and Kannan, J. Bharathi}
}

@book{Smith19,
  title     = {Modern Raman Spectroscopy: A Practical Approach},
  author    = {Ewen Smith and Geoffrey Dent},
  year      = {2019},
  publisher = {John Wiley \& Sons Ltd},
  isbn      = {9781119440550},
  doi       = {10.1002/9781119440598},
  url       = {https://doi.org/10.1002/9781119440598},
}

@book{Stoeckmann99,
  title     = {Quantum Chaos: An Introduction},
  author    = {St{\"o}ckmann, Hans-Jürgen},
  year      = {1999},
  publisher = {Cambridge University Press},
  address   = {Cambridge},
  isbn      = {978-0521592857},
}

@article{Tiwari24,
  title = {Dynamical localization and slow dynamics in quasiperiodically driven quantum systems},
  author = {Tiwari, V. and Bhakuni, D. S. and  Sharma, A.},
  journal = {Phys. Rev. B},
  volume = {109},
  issue = {16},
  pages = {L161104},
  numpages = {8},
  year = {2024},
  month = {Apr},
  publisher = {American Physical Society},
  doi = {10.1103/PhysRevB.109.L161104},
  url = {https://link.aps.org/doi/10.1103/PhysRevB.109.L161104}
}

@article{Dunlap86,
  title = {Dynamic localization of a charged particle moving under the influence of an electric field},
  author = {D. H. Dunlap and V. M. Kenkre},
  journal = {Phys. Rev. B},
  volume = {34},
  issue = {6},
  pages = {3625--3633},
  numpages = {0},
  year = {1986},
  month = {Sep},
  publisher = {American Physical Society},
  doi = {10.1103/PhysRevB.34.3625},
  url = {https://link.aps.org/doi/10.1103/PhysRevB.34.3625}
}

@article{Grossmann91,
  title = {Coherent destruction of tunneling},
  author = {F. Grossmann, T. Dittrich, P. Jung and P. H\"anggi},
  journal = {Phys. Rev. Lett.},
  volume = {67},
  issue = {4},
  pages = {516--519},
  numpages = {0},
  year = {1991},
  month = {Jul},
  publisher = {American Physical Society},
  doi = {10.1103/PhysRevLett.67.516},
  url = {https://link.aps.org/doi/10.1103/PhysRevLett.67.516}
}

@article{Das10,
  title = {Exotic freezing of response in a quantum many-body system},
  author = {A. Das},
  journal = {Phys. Rev. B},
  volume = {82},
  issue = {17},
  pages = {172402},
  numpages = {4},
  year = {2010},
  month = {Nov},
  publisher = {American Physical Society},
  doi = {10.1103/PhysRevB.82.172402},
  url = {https://link.aps.org/doi/10.1103/PhysRevB.82.172402}
}

@article{Haldar21,
  title = {Dynamical Freezing and Scar Points in Strongly Driven Floquet Matter: Resonance vs Emergent Conservation Laws},
  author = {Haldar, A.  and Sen, D. and Moessner, R. and   Das, A.},
  journal = {Phys. Rev. X},
  volume = {11},
  issue = {2},
  pages = {021008},
  numpages = {25},
  year = {2021},
  month = {Apr},
  publisher = {American Physical Society},
  doi = {10.1103/PhysRevX.11.021008},
  url = {https://link.aps.org/doi/10.1103/PhysRevX.11.021008}
}

@article{DAlessio13,
title = {Many-body energy localization transition in periodically driven systems},
journal = {Ann. Phys.},
volume = {333},
pages = {19-33},
year = {2013},
issn = {0003-4916},
doi = {https://doi.org/10.1016/j.aop.2013.02.011},
url = {https://www.sciencedirect.com/science/article/pii/S0003491613000389},
author = {L. D’Alessio and A. Polkovnikov}
}

@article{Billy08,
  title = {Direct observation of Anderson localization of matter waves in a controlled disorder},
  author = {Billy, J. and Josse, V. and Zuo, Z. and Bernard, A. and Hambrecht, B. and Lugan, P. and Cl\'ement, D. and Sanchez-Palencia, L. and Bouyer, P. and Aspect, A.},
  journal = {Nature},
  volume = {453},
  issue = {7197},
  pages = {891-894},
  year = {2008},
  doi = {10.1038/nature07000},
  url = {https://doi.org/10.1038/nature07000}
}

@article{Serbyn21,
  author    = {Serbyn, Maksym and Abanin, Dmitry A. and Papi{\'c}, Zlatko},
  title     = {Quantum many-body scars and weak breaking of ergodicity},
  journal   = {Nature Phys.},
  volume    = {17},
  number    = {6},
  pages     = {675--685},
  year      = {2021},
  doi       = {10.1038/s41567-021-01230-2},
  url       = {https://doi.org/10.1038/s41567-021-01230-2},
  issn      = {1745-2481}
}

@article{Turner18,
  author    = {Turner, C. J. and Michailidis, A. A. and Abanin, D. A. and Serbyn, M. and Papi{\'c}, Z.},
  title     = {Weak ergodicity breaking from quantum many-body scars},
  journal   = {Nature Phys.},
  volume    = {14},
  number    = {7},
  pages     = {745--749},
  year      = {2018},
  doi       = {10.1038/s41567-018-0137-5},
  url       = {https://doi.org/10.1038/s41567-018-0137-5},
  issn      = {1745-2481}
}

@article{Fishman89,
  title = {Scaling theory for the localization length of the kicked rotor},
  author = {S. Fishman and R. E. Prange and M. Griniasty},
  journal = {Phys. Rev. A},
  volume = {39},
  issue = {4},
  pages = {1628--1633},
  numpages = {0},
  year = {1989},
  month = {Feb},
  publisher = {American Physical Society},
  doi = {10.1103/PhysRevA.39.1628},
  url = {https://link.aps.org/doi/10.1103/PhysRevA.39.1628}
}

@book{Abramowitz65,
author = "Abramowitz, M. and Stegun, I.",
title  = "Handbook of Mathematical Functions",
publisher = {Dover Publications},
  address = {New York},
year = "1965" }

@book{NR07,
 author = {Press, William H. and Teukolsky, Saul A. and Vetterling, William T. and Flannery, Brian P.},
 title = {Numerical Recipes 3rd Edition: The Art of Scientific Computing},
 year = {2007},
 isbn = {0521880688, 9780521880688},
 edition = {3},
 publisher = {Cambridge University Press},
 address = {New York, NY, USA},
}

@book{Guckenheimer02,
  title={Nonlinear Oscillations, Dynamical Systems, and Bifurcations of Vector Fields},
  author={Guckenheimer, J. and Holmes, P.},
  isbn={9780387908199},
  lccn={70022436},
  issn={0066-5452},
  series={Applied Mathematical Sciences},
  url={https://books.google.es/books?id=cumGDmhaFnkC},
  year={2002},
  publisher={Springer},
 address = {New York},
}

@book{LL10,
   title = "Regular and chaotic dynamics",
   author = "Lichtenberg, Allan J. and Lieberman, M.~A.",
   series = "Applied Mathematical Sciences",
   publisher = "Springer",
   address = "New York, Berlin, Heidelberg",
   url = "http://opac.inria.fr/record=b1120113",
   isbn = "3-540-97745-7",
   year = "2010"
}

@article{Abal02,
  title = {Dynamical localization in quasiperiodic driven systems},
  author = {Abal, G. and Donangelo, R. and Romanelli, A. and Sicardi Schifino, A. C. and Siri, R.},
  journal = {Phys. Rev. E},
  volume = {65},
  issue = {4},
  pages = {046236},
  numpages = {7},
  year = {2002},
  month = {Apr},
  publisher = {American Physical Society},
  doi = {10.1103/PhysRevE.65.046236},
  url = {http://link.aps.org/doi/10.1103/PhysRevE.65.046236}
}

@article{Ammann98,
  title = {Quantum Delta-Kicked Rotor: Experimental Observation of Decoherence},
  author = {Ammann, H. and Gray, R. and Shvarchuck, I. and Christensen, N.},
  journal = {Phys. Rev. Lett.},
  volume = {80},
  issue = {19},
  pages = {4111--4115},
  numpages = {0},
  year = {1998},
  month = {May},
  publisher = {American Physical Society},
  doi = {10.1103/PhysRevLett.80.4111},
  url = {http://link.aps.org/doi/10.1103/PhysRevLett.80.4111}
}

@article{Anderson58,
  title = {Absence of Diffusion in Certain Random Lattices},
  author = {Anderson, P. W.},
  journal = {Phys. Rev.},
  volume = {109},
  issue = {5},
  pages = {1492--1505},
  numpages = {0},
  year = {1958},
  month = {Mar},
  publisher = {American Physical Society},
  doi = {10.1103/PhysRev.109.1492},
  url = {http://link.aps.org/doi/10.1103/PhysRev.109.1492}
}

@article{Casati79,
  author = {G. Casati and B. V. Chirikov and F. M. Izrailev and J. Ford}, 
  journal = {Lect. Notes Phys.},
  volume = {93},
  pages = {334},
  year = {1979},
  url = {https://link.springer.com/chapter/10.1007/BFb0021757}
}

@article{Chacon01,
title = "Chaotic behavior in a dissipative non-ideal periodically kicked rotator ",
journal = "Phys. Lett. A ",
volume = "281",
number = "4",
pages = "231 - 239",
year = "2001",
note = "",
issn = "0375-9601",
doi = "http://dx.doi.org/10.1016/S0375-9601(01)00134-7",
url = "http://www.sciencedirect.com/science/article/pii/S0375960101001347",
author = "R. Chac\'on and A. Mart\'inez Garc\'ia-Hoz"
}

@article{Chacon09,
  author={R. Chac\'on and F. Borondo and D. Farrelly},
  title={Controlling dynamical localization by waveform reshaping},
  journal={EPL (Europhysics Letters)},
  volume={86},
  number={3},
  pages={30004},
  url={http://stacks.iop.org/0295-5075/86/i=3/a=30004},
  year={2009}
}

@article{Fishman82,
  title = {Chaos, Quantum Recurrences, and Anderson Localization},
  author = {Fishman, Shmuel and Grempel, D. R. and Prange, R. E.},
  journal = {Phys. Rev. Lett.},
  volume = {49},
  issue = {8},
  pages = {509--512},
  numpages = {0},
  year = {1982},
  month = {Aug},
  publisher = {American Physical Society},
  doi = {10.1103/PhysRevLett.49.509},
  url = {http://link.aps.org/doi/10.1103/PhysRevLett.49.509}
}

@article{Grempel84,
  title = {Quantum dynamics of a nonintegrable system},
  author = {Grempel, D. R. and Prange, R. E. and Fishman, Shmuel},
  journal = {Phys. Rev. A},
  volume = {29},
  issue = {4},
  pages = {1639--1647},
  numpages = {0},
  year = {1984},
  month = {Apr},
  publisher = {American Physical Society},
  doi = {10.1103/PhysRevA.29.1639},
  url = {https://link.aps.org/doi/10.1103/PhysRevA.29.1639}
}

@article{Hamilton22,
  title = {Classical-quantum localization in one dimensional systems: The kicked rotor},
  author = {Hamilton, C. and Pérez-Ríos, J.},
  journal = {AIP Advances},
  volume = {12},
  number = {3},
  pages = {035040},
  year = {2022},
  doi = {10.1063/5.0084028},
  publisher = {American Institute of Physics}
}

@article{Lignier05,
  author={H. Lignier and J. C. Garreau and P. Szriftgiser and D. Delande},
  title={Quantum diffusion in the quasiperiodic kicked rotor},
  journal={Europhys. Lett.},
  volume={69},
  number={3},
  pages={327},
  url={http://stacks.iop.org/0295-5075/69/i=3/a=327},
  year={2005},
}

@article{Medhet20,
  author    = {Sara Medhet and Tomotake Yamakoshi and Muhammad Ayub and Farhan Saif and Shinichi Watanabe},
  title     = {Signatures of inter-band transitions on dynamical localization},
  journal   = {The European Physical Journal D},
  volume    = {74},
  number    = {8},
  pages     = {175},
  year      = {2020},
  doi       = {10.1140/epjd/e2020-100489-1},
  issn      = {1434-6079}
}

@article{Moore94,
  title = {Observation of Dynamical Localization in Atomic Momentum Transfer: A New Testing Ground for Quantum Chaos},
  author = {Moore, F. L. and Robinson, J. C. and Bharucha, C. and Williams, P. E. and Raizen, M. G.},
  journal = {Phys. Rev. Lett.},
  volume = {73},
  issue = {22},
  pages = {2974--2977},
  numpages = {0},
  year = {1994},
  month = {Nov},
  publisher = {American Physical Society},
  doi = {10.1103/PhysRevLett.73.2974},
  url = {http://link.aps.org/doi/10.1103/PhysRevLett.73.2974}
}

@article{Revuelta15,
  author={F. Revuelta and R. Chac{\'o}n and F. Borondo},
  title={Towards AC-induced optimum control of dynamical localization},
  journal={EPL (Europhysics Letters)},
  volume={110},
  number={4},
  pages={40007},
  url={http://stacks.iop.org/0295-5075/110/i=4/a=40007},
  year={2015}
}

@article{Revuelta18,
  title = {Dynamical localization in non-ideal kicked rotors},
  author = {F. Revuelta and R. Chac{\'o}n and F. Borondo},
  journal = {Phys. Rev. E},
  volume = {98},
  issue = {12},
  pages = {062202},
  numpages = {7},
  year = {2018},
  month = {Dec},
  publisher = {American Physical Society},
  doi = {10.1103/PhysRevE.98.062202},
  url = {https://doi.org/10.1103/PhysRevE.98.062202}
}

@article{Ringot00,
  title = {Experimental Evidence of Dynamical Localization and Delocalization in a Quasiperiodic Driven System},
  author = {Ringot, J. and Szriftgiser, P. and Garreau, J. C. and Delande, D.},
  journal = {Phys. Rev. Lett.},
  volume = {85},
  issue = {13},
  pages = {2741--2744},
  numpages = {0},
  year = {2000},
  month = {Sep},
  publisher = {American Physical Society},
  doi = {10.1103/PhysRevLett.85.2741},
  url = {http://link.aps.org/doi/10.1103/PhysRevLett.85.2741}
}

@article{Steck00,
  title = {Quantitative study of amplitude noise effects on dynamical localization},
  author = {Steck, Daniel A. and Milner, Valery and Oskay, Windell H. and Raizen, Mark G.},
  journal = {Phys. Rev. E},
  volume = {62},
  issue = {3},
  pages = {3461--3475},
  numpages = {0},
  year = {2000},
  month = {Sep},
  publisher = {American Physical Society},
  doi = {10.1103/PhysRevE.62.3461},
  url = {https://link.aps.org/doi/10.1103/PhysRevE.62.3461}
}

\end{document}